\documentclass[journal=jacsat,manuscript=article]{achemso}

\usepackage[version=3]{mhchem} % Formula subscripts using \ce{}

\def\be{\begin{equation}}
\def\ee{\end{equation}}
\def\bea{\begin{eqnarray}}
\def\eea{\end{eqnarray}}
\def\a{\alpha}

\author{Anirban Polley$^1$}
\author{Satyavani Vemparala$^2$}
\author{Madan Rao$^{1,3}$}
\email{madan@rri.res.in}
\affiliation[University]
{$^1$Raman Research Institute, C.V. Raman Avenue, Bangalore 560080,
India\\
$^2$The Institute of Mathematical Sciences, CIT Campus, Chennai 600 113, India\\
$^3$National Centre for Biological Sciences (TIFR), Bellary Road,
Bangalore 560065, India}

\title[\texttt{achemso} demonstration]
{Atomistic simulations of a multicomponent asymmetric lipid bilayer}

\begin{document}
%%%%%%%%%%%%%%%%%%%%%%%%%%%%%%%%%%%%%%%%%%%%%%%%%%%%%%%%%%%%%%%%%%%%%
%% The manuscript does not need to include \maketitle, which is
%% executed automatically.  The document should begin with an
%% abstract, if appropriate.  If one is given and should not be, the
%% contents will be gobbled.
%%%%%%%%%%%%%%%%%%%%%%%%%%%%%%%%%%%%%%%%%%%%%%%%%%%%%%%%%%%%%%%%%%%%%
\begin{abstract}
The cell membrane is inherently asymmetric and heterogeneous in its composition, a feature that is crucial for its function. 
Using atomistic molecular dynamics simulations, the physical properties of a 3-component asymmetric mixed lipid bilayer system 
comprising of an unsaturated POPC (palmitoyl-oleoyl-phosphatidyl-choline), a saturated SM (sphingomyelin) and cholesterol are investigated.
In these simulations, the initial stages of  liquid ordered, $l_o$, domain formation are observed and such domains are found to be highly 
enriched in cholesterol and SM. The current simulations also suggest that the cholesterol molecules may partition into these SM-dominated regions 
in the ratio of $3:1$ when compared to POPC-dominated regions. SM molecules exhibit a measurable tilt and long range tilt correlations are observed 
within the $l_o$ domain as a consequence of the asymmetry of the bilayer, with implications to local membrane deformation and budding. Tagged particle 
diffusion for SM and cholesterol molecules, which reflects spatial variations in the physical environment encountered by the tagged particle, is computed 
and compared with recent experimental results obtained from high resolution microscopy.
\end{abstract}

%%%%%%%%%%%%%%%%%%%%%%%%%%%%%%%%%%%%%%%%%%%%%%%%%%%%%%%%%%%%%%%%%%%%%
%% Start the main part of the manuscript here.
%%%%%%%%%%%%%%%%%%%%%%%%%%%%%%%%%%%%%%%%%%%%%%%%%%%%%%%%%%%%%%%%%%%%%
\section*{Introduction}
The cell membrane is characterized by both lateral and transverse lipid heterogeneity, an aspect of significant functional consequence \cite{devaux}.
Transverse lipid heterogeneity is maintained actively by the cell, making the cell bilayer intrinsically asymmetric. Lateral lipid 
heterogeneities called `rafts' \cite{simons1,simons2}, which are ternary mixtures of sphingomyelin (SM), phosphotidyl Choline (PC) and cholesterol 
(Chol) molecules have been implicated in a variety of cellular processes including signaling and endocytosis, even though 
 the nature of these functional cellular rafts is still a matter of contention \cite{TrafficRaoMayor,raftreviews,sharma,debanjan}. Asymmetry in 
bilayers can arise both in terms of difference in constituent lipids or in number of lipid molecules in both leaflets. In spite of this obvious lateral/transverse 
compositional heterogeneity, except for a few seminal studies\cite{kellerAS}, most in-vitro investigations of multicomponent artificial membranes  
\cite{keller,keller03,webb}  have been done on symmetric bilayers. Further,  most of the atomistic simulations of model cell membrane mimics, 
have been carried out on systems which either have lateral heterogeneity \cite{mikko} or transverse asymmetry  \cite{Martin,Andrey,AndreyA} but rarely both.
There have however been a few studies using coarse-grained simulations \cite{laradji,wagner} and continuum Landau theories \cite{hansen,olmsted,schick}
that  address the role of inter-bilayer coupling in equilibrium physical properties and domain growth.

While early simulations of model membranes consisted only of a single-component PC bilayers \cite{Berendsen94,Berger97,Tieleman98}, later simulations have incorporated more than one lipid component, in particular cholesterol \cite{Holtje01}. Following the `raft proposal'  of the importance of SM molecules in raft formation, an increasing number of simulations with SM molecules \cite{Vattulainen04} have been carried out. These simulations include detailed comparisons between mixtures of SM and Chol 
%have been carried out 
%ave shown that due to additional hydrogen bonding
%capabilities of SM molecules, cholesterol molecules have stronger affinity to SM molecules compared to 
and PC and Chol \cite{mikko,aitto,ohvo,rog,george,zhang}. 
%However, a simulation study by Zhang etal suggested that energetically there is no difference between interactions of cholesterol with PC or SM molecules
%and that a subtle interplay of energetics and entropy may play a significant role in domain formation observed where $l_o$ domains are
%enriched by SM and Chol molecules. 
Finally, simulations on asymmetric bilayers have also been reported. A simulation study by Bhide et al. \cite{bhide} was 
performed on systems consisting of SM and Chol molecules in upper leaflet and SOPS and Chol molecules in the lower leaflet, with equal number of 
phospholipids in both the leaflets. Comparisons between each leaflet of the asymmetric bilayer with corresponding simulations of symmetric bilayers
showed no significant differences in their physical properties.
% when compared with results of symmetric bilayers, suggesting almost non-existing 
%cross-leaflet interactions.

In this paper, the equilibrium properties of a multicomponent asymmetric (both lateral and transverse) bilayer using atomistic molecular dynamics (MD) simulations are 
studied. Specifically, the physical properties of a 3-component asymmetric lipid bilayer comprising of an unsaturated POPC, a saturated 
SM and Chol molecules, which exhibits lateral compositional heterogeneities in the form of liquid ordered ($l_o$) - liquid disordered ($l_d$) domains,
thought to represent the characteristic lipid composition of rafts on the cell membrane. 
%In addition to studying a variety of 
%order parameters, their correlations and spatial distribution, we have studied transport properties of the components. Our study shows that the presence of lateral heterogeneities in the bilayer affect molecular diffusion at short time scales; this may be of relevance to molecular diffusion at the cell surface.
In addition to studying a variety of order parameters, their correlations and spatial distribution, transport properties of the component lipid molecules are also studied. This study suggests that the presence of lateral heterogeneities in the bilayer can potentially affect the molecular diffusion at short time scales, which may be of relevance to molecular diffusion at the cell surface.
 However lipid based lateral heterogeneities on the cell surface are likely to be small and below optical resolution; it is only recently that 
advances in high resolution single-particle tracking (SPT) \cite{Kusumi05,hell} allow one to measure subtle changes in diffusion characteristics 
as molecules move across the heterogeneous cell surface. In addition, the transverse asymmetry would suggest that the diffusion characteristics of the molecule 
is different in the two leaflets of the bilayer. Correlations between the local physical environment and the tagged particle diffusion are also investigated in this
paper. 

\section*{Materials and Methods}
Two bilayer model systems have been simulated : ($A$) a symmetric three-component bilayer consisting of $170$ POPC, $171$ palmitoyl-sphingomyelin 
(SM, saturated lipid) and $171$ Chol (ratio of $1:1:1$) in each leaflet and ($B$) an asymmetric bilayer 
%with a slightly different number of lipids in the two 
%leaflets and 
with a composition of $170$ POPC, $168$ SM and $171$ Chol (ratio of roughly $1:1:1$) in the upper leaflet and $256$ POPC and $256$ Chol 
(ratio $1:1$)in the lower leaflet. Two control systems have also been simulated, which are ($C$) a symmetric one-component bilayer made up of $128$ 
pure POPC and ($D$) a symmetric two-component bilayer made up of $64$ POPC and $64$ Chol (ratio $1:1$ in both leaflets).

%(in upper leaflet) asymmetric bilayer with the same composition as in ($B$) 

The force field parameters for POPC were taken from the previously validated united-atom description (Tielman and Berendsen \cite{Tieleman-POPC}), 
whereas for SM and Chol, parameters from works of  Niemela et al.\cite{mikko} were used. Water was simulated by the simple point charge (SPC) model  
\cite{Berendsen} and PACKMOL \cite{packmol} was used to generate the initial configurations of all the bilayer systems and hydrated 
with water in the ratio of 1(lipid) : 32 (water).

All MD simulations were performed at a temperature $T=296$K, which lies between the main transition temperatures of POPC ($T_m= -2.9\pm 1.3^\circ$C) 
\cite{koynova98} and SM ($T_m\approx41.4^\circ$C) \cite{koynova95} and right in the middle of the $l_o$-$l_d$ phase coexistence region of the symmetric 
ternary system \cite{prieto}. GROMACS (http://www.gromacs.org) \cite{Lindahl} software was used to integrate the equations of motion with a time step 
of $2$\,fs. To avoid bad contacts arising from steric constraints during initialization, all bilayer systems were subjected to steepest descent 
minimization initially. The systems were then simulated for $50$\,ps in the NVT ensemble using a Langevin thermostat. Subsequently, each system was 
simulated in the NPT ensemble ($T = 296$K, $P =1$atm) using a Berendsen thermostat and semi-isotropic pressure coupling with compressibility 
$4.5\times 10^{-5}$ bar$^{-1}$  for $220$\,ns (for the asymmetric ternary system $B$) and for $100$\,ns (for symmetric ternary system $A$) and 
for $100$\,ns (for the single and two -component systems $C$ and $D$). Lipid system parameters such as deuterium order parameter and mean thickness were monitored 
throughout the simulations to ensure that the system is well equilibrated.  All the reported results are computed over last $20$ns  unless otherwise 
stated.

Note that the two model systems $A$ and $B$ have been simulated starting from two sets of initial conditions : (i) where the components in each leaflet are mixed or (ii) where the ternary components are completely phase segregated.  

\section*{Results}

To make sure that a stable, surface tension-less asymmetric bilayer is being simulated, the forces, torques and surface tension of the bilayer are
computed from the local stress tensor $\sigma_{ij}(x,y,z)=\frac{1}{v}\sum_{\a}f_{i}^{\a}r_{j}^{\a}$, where $f_i^{\a}$ is 
the $i^{th}$ component of the force on the $\a^{th}$-particle due to all other particles within a coarse-grained 
volume $v=[0.1\,nm]^3$. 
%%%%%%%%%%%%%%%%%%%%%%%%%%%%%
% We computed the net force 
%  $F_i =\int \partial_{k}\sigma_{ik} \,dv$
%  and its first moment (related to the torque) $M_{ik}=\int(\partial_l \sigma_{il} x_{k}-\partial_l\sigma_{kl} x_{i})\,dv$, and  find that at equilibrium we achieve both
%  force and torque balance -- $F_1=  0.29\pm 3.34$ , $F_2 = -2.56 \pm 2.09$, $F_3 = -2.16\pm 3.2$, in units of nN and   
%  $M_{12}=0.77\pm0.428, M_{13}=-8.83\pm2.04, M_{23}=-13.303\pm1.67$, in units of nN$\cdot$nm. These values are comparable to the
%  corresponding values for the symmetric bilayers (Supplementary Information) implying that our model asymmetric bilayer is mechanically stable.
%Integrating over the plane of the membrane gives
% the stress tensor at each transverse slice of the bilayer, $\overline{\sigma}_{ij}(z)$.
The net force $F_i =\int \partial_{k}\sigma_{ik} \,dv$ and its first moment (related to the torque) $M_{ik}=\int(\partial_l \sigma_{il} x_{k}-\partial_l\sigma_{kl} x_{i})\,dv$ for the bilayer were computed and both force and torque balance -- $F_1=  0.29\pm 3.34$ , $F_2 = -2.56 \pm 2.09$, $F_3 = -2.16\pm 3.2$, in units of nN and $M_{12}=0.77\pm0.428, M_{13}=-8.83\pm2.04, M_{23}=-13.303\pm1.67$, in units of nN$\cdot$nm were achieved suggesting a mechanically stable asymmetric bilayer. These are comparable to the corresponding values for the symmetric bilayers (see Supplementary Information).
%%%%%%%%%%%%%%%%%%%%%%%%%%%
The membrane surface tension is calculated  as $\gamma=\int\pi(z)dz$, integrated over the width of the 
bilayer, where $\pi(z)$ is the lateral pressure, given by  $\pi(z)=\frac{1}{2} \left({\overline \sigma}_{xx}(z)+
{\overline \sigma}_{yy}(z)\right)-{\overline \sigma}_{zz}(z)$ \cite{safran,patra}, resulting in a value of $\gamma=-0.0018\pm0.0301$\,
bar-nm, essentially a `zero' surface tension bilayer. More details of time dependence of net forces, their moments and 
surface tension are provided in Supplementary Figure S2. A snapshot of the equilibrium bilayer configuration of the 
asymmetric membrane %(\,\ref{MD_snap})
 is shown in Figure 1(a). In  Figure 1(b), the pressure profiles of symmetric and 
asymmetric bilayers are shown for comparison. In contrast to the asymmetric bilayer, the symmetric bilayer exhibits a symmetric pressure 
profile about the midplane. On the other hand the pressure profile of the asymmetric bilayer shows larger spatial oscillations.

%The profiles for the upper leaflet ($z>0$) of the two bilayers is similar due to
%near identical composition of the lipids. However the pressure profile for the lower leaflet differs between the symmetric and asymmetric
%bilayers due to higher amount of cholesterol present in the lower leaflet in the case of asymmetric bilayer($50\%$). The pressure gradient increases 
%due to the high elastic modulus of the lower leaflet as a consequence of high rigidity in the bilayer due to presence of high 
%concentration of cholesterol as seen in earlier simulations as well \cite{patra}.

\subsection*{Lipid composition}
At the temperature and overall lipid composition under consideration, the homogeneous mixed phase of both the asymmetric and symmetric bilayers 
is unstable and exhibits definite features of phase separation between POPC-rich and SM-rich domains (Supplementary Figures, S3, S4); in the asymmetric bilayer 
phase segregation occurs in the upper leaflet alone, while the composition in the lower leaflet remains homogeneous. To show 
that the bilayer is undergoing phase separation towards a final equilibrium phase segregated configuration,
%signatures of initial-state domain formation in the asymmetric bilayers, 
the theory of dynamical coarsening \cite{Bray}, which 
deals with the study of the dynamics of domain formation starting from a complete disordered phase
% before reaching an equilibrium state, 
is used.  
%In this work, signature of phase separation is given by evaluating the 
%time dependence of energy density  $\epsilon = E/V$, where the energy $E \propto \int d^{2}r \left(\nabla\phi\right)^{2}$ and 
%order parameter is defined as $\phi=\frac{\rho_{\tiny{\mbox{SM}}}-\rho_{\tiny{\mbox{POPC}}}}{\rho_{\tiny{\mbox{SM}}}+\rho_{\tiny
%{\mbox{POPC}}}}$. 
In simulations starting from an initial mixed state, it was found that the system quickly phase segregates;
at early times the domain sizes are small and grow over the simulation time, $220$\,ns. It has been shown earlier that complete 
phase segregation occurs only over a very long time, at least on a microsecond time scale \cite{pandit,marrink}. To demonstrate 
dynamics of domain formation or coarsening, we have computed 
%In this study, 
%the coarsening of domains is demonstrated by  computing 
(i) the probability distribution $P(\phi)$ of the order parameter $\phi=\frac{\rho_{\tiny{\mbox{SM}}}-\rho_{\tiny{\mbox{POPC}}}}{\rho_{\tiny{\mbox{SM}}}+\rho_{\tiny
{\mbox{POPC}}}}$. 
at different times (where $\rho_{\tiny{\mbox{SM/POPC}}}$ is the local density of SM/POPC),
and (ii) the energy density  $\epsilon = E/V$, where the energy $E \propto \int d^{2}r \left(\nabla\phi\right)^{2}$,
which measures the time dependence of the amount of interface separating the two phases. The results are shown in Figure 2. 
Initially $P(\phi)$ is peaked at $\phi=0$ (mixed state) and subsequently evolves into a distribution with two peaks at $\pm 1$ 
which gets progressively sharper with time. Dynamical scaling in the coarsening regime \cite{Bray} implies that  
$\epsilon \sim t^{-1/z}$, where $z=3$, since the order parameter is conserved. 
Our simulations show that $1/z=0.30\pm0.15$, consistent with this scaling prediction (Figure 2 (a)).

The domains are observed to be larger in the symmetric bilayers compared to the asymmetric bilayer, when measured over the total simulation 
time scale ($220$\,ns). The interfacial fluctuations of the $l_o$-domains in the asymmetric bilayer are larger than the symmetric bilayer, 
suggesting, in line with earlier simulations \cite{marrink}, that the interfacial tension between the domains is higher in the symmetric bilayer.
% may be the 
%cause for such fluctuations. 
Both the larger interfacial tension and the enhanced correlations between the domains in the two leaflets of the symmetric bilayer
(seen in Supplementary Figure, S4), are a result of a transbilayer coupling. This transbilayer correlation between domains can be understood 
within a Landau description of both equilibrium and dynamics, in which the relative concentration of lipids in one leaflet acts as a  
``field'' for the concentration in the other leaflet \cite{schick}. Local equilibrium relates the domain growth velocity to the
 interfacial tension \cite{Bray}, this suggests that the domains coarsen faster in the symmetric bilayer, consistent with our simulations.

To estimate the relative partitioning of cholesterol in the POPC and the SM rich regions in the upper leaflet, the joint probability 
distribution of finding a given concentration of SM with Chol in an $xy$ region (similarly, POPC with Chol) is calculated
and shown in Figure 2 (b) and 2 (c). These joint probabilities  show that the cholesterol concentration completely correlates with
the SM concentration (i.e., Chol is low (high) when the SM concentration is low (high)) and completely {\em anticorrelates} with 
the POPC concentration (i.e., Chol is low (high) when POPC is high (low)). The joint probability densities plotted in Figures 2(b)
and (c) strongly suggest that cholesterol preferentially partitions in the SM-rich phase three times more than in POPC-rich
region (more precisely $2.97:1$)

The segregation of chemical composition is accompanied by changes in physical characteristics of the segregated molecules. 
%The saturated lipid tails of SM that are in the SM-enriched domains are found to be more rigid than both the SMs in the POPC-rich domains and POPC.
The saturated lipid tails of SM that are in the SM-enriched domains are found to be more rigid than both the SM and POPC molecules in the POPC-rich domains.
 The deuterium order parameter ($S$) describing the rigidity of lipid tails is computed for SM molecules in both
the leaflets and is shown in Figure 4. The probability distribution, $P(S)$, of SM gives indication of the location of SM in either
SM-rich (higher value) or POPC-rich domains (lower value). As seen in Figure 3(a), $P(S)$ has a bimodal distribution. The distinction
of the deuterium order parameter in the two regions indicate that the SM- and POPC-rich regions may be identified with $l_o$-like and $l_d$-like phases,
respectively. 
From Figure 3(a), it can be seen that the probability distribution of deuterium order parameters in the asymmetric bilayers are 
comparable to that of symmetric bilayer. The spatial variation of the deuterium order parameter in the $x-y$ plane for both 
asymmetric and symmetric bilayers is shown in Supplementary Figure S7 and it can be noted that the segregation of chemical composition 
is naturally accompanied by an $l_{o}$-$l_{d}$-like phase 
separation in the bilayer membrane. From this spatial distribution,
it can be seen that the size of $l_o$-like domain size (enriched in SM) is significantly larger in the symmetric bilayer compared to 
that of asymmetric bilayer, over the time scale measured.

The probability distribution and the spatial profile of the local bilayer thickness is also computed and shown in 
Figure 3(b) and Supplementary Figure S7 respectively. The SM-enriched $l_o$-like domains have a larger bilayer thickness compared 
to the POPC- enriched $l_d$-like domains. As with deuterium order parameter, the distribution of bilayer thickness for both the 
symmetric and asymmetric bilayers is bimodal, corresponding to the $l_d$-like and $l_o$-like domains. It is significant that the 
difference in the bilayer thickness between the $l_o$-like and $l_d$-like phases in the symmetric bilayer is consistent with 
recent AFM studies \cite{Kruija}, and is {\em more than twice compared to asymmetric bilayer}. The spatial profile of the thickness shown 
in Supplementary Figure S7 is an accurate measure of the domain size and consistent with the deuterium order parameter results, the 
domain size in the symmetric bilayer is larger than that of asymmetric bilayer over the time scale measured. 

By computing the joint probability of finding SM molecules in the upper and lower leaflets of the symmetric bilayer, it can be seen 
(Supplementary Figure S5) that there is a clear registry of SM-rich ($l_o$-like) domains across the leaflets, which may be   
responsible for observed increase in SM-rich domain sizes in symmetric bilayers.

\subsection*{Lipid splay and tilt}
To quantify the relative packing of lipid chains, the amount of splay between the two lipid tails is calculated, in addition to 
deuterium order parameter computed above. A tail vector is defined as a vector originating from the carbon of the carbonyl group and 
pointing to the terminal methyl carbon of a lipid tail.  The splay angle is the angle between two such tail vectors of each 
lipid. Simulation results suggest that the extent of splay in SM lipid tails is significantly smaller in $l_o$-like domain 
($17^\circ$) compared to POPC-rich domain ($40^\circ$) as seen in Supplementary Figure S6. In addition, the extent of lipid 
splay in the symmetric and asymmetric bilayers is comparable. The spatial heterogeneity of the splay in the asymmetric bilayer 
reflects the differences in lipid composition (Supplementary Figure, S8).  Taken together with the deuterium order parameter data, 
the lipid splay results strongly suggest that the packing fraction of lipids in the SM-rich phase is higher than that in 
the POPC-rich phase.

Tilt angle of the lipid tail chain is defined as the orientation of the mean tail vector of the lipid with respect to the local 
outward normal to the membrane, and is described by two angles $(\theta, \phi)$, the polar and azimuthal angles, respectively. 
The angle $\phi$ measures the orientation of the 2d tilt vector, the projection of the tail vector onto the tangent plane, 
with the $x$-axis. An accurate determination of  the  tilt angles of the component lipids is quite involved, since, over short 
length scales, the local normal to the membrane fluctuates due to molecular protrusion effects. To compute local average tilt,
a coarse-graining scale is chosen, which should be more than the protrusion scale and less than the tilt 
correlation length (which in turn should of course be smaller than the size of the $l_o$-like domain). Over this coarse-grained
scale, a membrane normal is considered to be along the $z$-axis. A convenient choice of coarse graining scale is around a  
$1$\,nm$^2$ (which encompasses $\approx 3$ lipids on an average), for which statistically reliable results can be obtained.
For instance, for POPC-only bilayer, the probability distribution of the coarse-grained angle, $P(\theta)$,  is peaked about 
zero, while the distribution $P(\phi)$ is uniform, consistent with the known fact that POPC does not exhibit a tilt at this 
temperature. 

Following such a procedure for computing local tilt angles for the ternary membrane is problematic, since over the timescale of 
the MD simulation, the size of the $l_o$-like domains are small, making it difficult to obtain an unambiguous  determination of 
tilt and its correlations. 
%To address this issue, simulation of a ternary bilayer starting from a completely 
%phase segregated configuration with cholesterol distributed, between SM-rich and POPC-rich domains, in the ratio of 3:1
%(as described in Materials and Methods).
To address this issue, simulation of a ternary bilayer starting from a completely phase segregated configura-
tion with cholesterol distributed, between SM-rich and POPC-rich domains, in the ratio of 3:1 (as
described in Materials and Methods) is carried out.
 In this case the $l_o$-like domain (SM-rich) size is around half the system size and can
 provide good statistics. 

The tilt angle distribution of POPC lipids in the symmetric and asymmetric ternary bilayers has a similar distribution as in the 
case of POPC-only bilayer as seen in Supplementary Figure S9. On the other hand, the tilt angle distribution for SM shows an 
interesting trend. While SM in the symmetric bilayer shows no evidence of tilt ($P(\theta)$ is peaked at around $0^{\circ}$ and 
$P(\phi)$ is uniform), SM in the SM-rich ($l_o$-like) domain of the asymmetric bilayer has a nonzero tilt of around $2.016 \pm 
0.69^{\circ}$. This small tilt of SM in the asymmetric bilayer is consistent with the decrease in bilayer thickness of the 
asymmetric membrane compared to that of symmetric bilayer (see Figure 3(b)). Tilt angle correlations defined as $C(r) = \langle 
\phi(r) \phi(0)\rangle$ are computed for POPC and SM lipid molecules and the results are shown in Figure 4. The correlation 
functions of the tilt of POPC and SM (symmetric bilayer) decay exponentially to zero, consistent with the above findings. 
However, $C(r)$ for SM in the asymmetric bilayer decays exponentially to a non zero value, signaling long range order 
\cite{Chaikin-Lubensky} in SM molecules in the asymmetric bilayer. This suggests that the asymmetric nature of the bilayer 
can potentially generate a tilt ordering of SM molecules in the SM-rich ($l_o$-like) domain.

Previous simulations \cite{marrink,schafer,mikko} and experiments \cite{rinia,soni,keller,garcia,brown,saslowski} have shown that the 
thickness of $l_o$ domains is larger than that of $l_d$ domain. This difference in thickness between $l_o$ and $l_d$ domains gives rise 
to line tension along the domain boundary \cite{garcia}. The lipid tails of the $l_o$ domain can be exposed to solvent as a result of 
such mismatch, which is energetically unfavourable. One of the ways to mitigate such mismatch is for the thicker $l_o$ domain to undergo 
a small tilt such that the head groups of the two domains can be at the same height. 
%The observed small tilt in the simulations of 
%asymmetric bilayers support this hypothesis.
The observed small tilt in the simulations of asymmetric bilayers supports this hypothesis.
 Within a Landau theory, the asymmetric bilayer can be thought of as being subjected 
to a transverse compression, which would naturally lead to a tilt when the lipids are stretched out (as they are in the $l_0$ domain)
 \cite{Arif,Arif1}.

The existence of a finite tilt and its correlation over long scales, if verified experimentally \cite{Tonya}, could have important consequences 
for membrane deformation and budding \cite{sarasij}. The tilt vector naturally couples to the local curvature tensor of the 
membrane, giving rise to anisotropic bending stresses. If the constituent molecules are chiral (as they usually are), then there 
are additional bending stresses coming from  chiral couplings of the tilt and curvature.  If strong enough, these bending stresses 
can induce membrane deformation giving rise to spherical buds or cylindrical tubules \cite{sarasij,garcia}.

\subsection*{Tagged particle diffusion}
Even though the membrane composition forms stable domains at equilibrium, the individual component molecules can traverse across 
domains. The equilibrium dynamics of certain `tagged' component lipids can be monitored by measuring their 
diffusion coefficients and correlating them with the physical and chemical heterogeneity across the membrane. The mean square 
displacement (MSD) is defined as $\langle \delta r_i(t)^2 \rangle$ of a tagged particle, where  $\delta r_i(t)=r_i(t)-r_i(0)$ is the 
displacement of tagged $i^{th}$ lipid of a given species at time $t$ from its position at $t=0$. While computing the 
MSD, the location of the tagged lipid, whether it is in the POPC-rich ($l_d$-like) or SM-rich ($l_o$-like) domain, is monitored
by computing the instantaneous deuterium order parameter $S$ of the tagged lipid and the local bilayer thickness. The diffusion
analysis is show in Figure 5. It can be seen that the diffusion coefficient of both SM and Chol molecules, given by the 
saturation value of  $\langle \delta r_i^2(t) \rangle/4t$, is different depending on whether the tagged lipid molecule is 
in the $l_o$-like and $l_d$-like domain as has been reported in other simulation studies\cite{marrink}.

For tagged SM or Chol molecules that initially lie in the POPC-rich ($l_d$-like) domain, the local diffusion coefficient
crosses over from an early time high value ($D_0$) to a late time low value ($D_{\infty}$). For each of the tagged components, 
a first-passage  time, defined as the first time that a tagged molecule residing in a domain moves out of it, is computed. The 
crossover time $\tau$ is then obtained from such a computed first-passage time. The MSD is found to obey a crossover scaling relation,
\be
\langle \delta r^2 \rangle =  4 D_0 t F\left( t/\tau \right) 
\ee
where the nonlinear scaling function $F$ has the asymptotic form,
\bea
F\left( t/\tau \right)&=& 1 \,\,\,\,\,\,\,\,\,\,\,\,\,\,\,\, \mbox{for} \,\,\,\,\,\,\ t/\tau \ll  1 \nonumber \\
                              &=& \frac{D_{\infty}}{D_0}  \,\,\,\,\,\,\, \mbox{for}\,\,\,\,\,\,\,\,  t/\tau \gg 1 
\eea
%$\langle \delta r^2 \rangle =  4 D_0 t F\left( t/\tau \right)$
%where the nonlinear scaling function $F$ has the asymptotic form,
%$F\left( t/\tau \right) = 1 $ \, \mbox{for} $t/\tau \ll  1$, and
%$F\left( t/\tau \right)=\frac{D_{\infty}}{D_0} $ \, \mbox{for} $t/\tau \gg 1$.

The data collapse shown in Figures 5(a) and 5(b) demonstrate this crossover scaling for SM and Chol molecules, respectively.
Experimentally the crossover time scale $\tau$ can be obtained from the value of $t$ at which the instantaneous $D(t) =
\left(D_0 + D_{\infty}\right)/2$ , rather than the first-passage time.%{\bf [references to be included here for this statement]} 
The values of the diffusion coefficients for SM molecules (Fig. 5(a))
%{\bf [these numbers both from simulation and experiment should be given instead of just stating it.]}
 are in agreement with the experimental findings \cite{Filippov,Pralle}, while the values for Chol 
molecules (Fig. 5(b)) are slightly lower than those reported in other simulations \cite{marrink}. This discrepancy can be attributed to the 
differences in lipid composition of the ternary bilayer used in both the simulations.
The difference in the tagged particle diffusion coefficient between the two domains can be attributed to changes in local 
viscosity ($\eta$), moment-of-inertia ($I$) of the particle and to changes in the local density correlations (given by the 
local partial structure factors, $S_{\alpha \beta}(q)$) with neighbouring molecules that the tagged particle experiences 
as it traverses across the domain. The change in the deuterium order parameter $S$ is used to track the changes in 
moment-of-inertia of the tagged molecules. For SM molecules, this shows a crossover similar to the crossover diffusion, 
with the value of $S$ being low (high) when the diffusion coefficient is high (low) as seen in Figure 5(a). On the other 
hand cholesterol being a rigid molecule is unlikely to undergo any conformational change, resulting change in moment-of-inertia,
as it traverses across the domains. Thus a substantial  change in the diffusion coefficient of Chol molecules can be attributed 
to the changes in viscosity and local density correlations, $S_{\alpha \beta}(q)$, arising from changes in the local environment.

In the context of SPT experiments, if the time scale over which the tagged particle is tracked is large enough so that the 
particle crosses and recrosses the domains, then one would  measure the MSD of $\delta r_i(t)= \int dt^{\prime} \left(r_i(t^
{\prime}+t)-r_i(t^{\prime})\right)$. For the control pure POPC symmetric bilayer, it is seen that the trajectories are purely 
diffusive ($\langle \delta r_i(t)^2 \rangle \propto t$) in nature. For the ternary bilayer, however, this MSD would exhibit 
deviations from true diffusion, which can be characterized by $\langle \delta r_i(t)^2 \rangle \propto t^{\alpha}$. 
Figure 5c shows the MSD for the tagged lipids in both the symmetric and asymmetric bilayers and the corresponding values of the
exponent $\alpha$. The value of $\alpha$ obtained for SM in the asymmetric bilayer ($\alpha\approx 0.39$) is consistent with 
the experimental value obtained by analyzing recent SPT of labeled SM on the plasma membrane of epithelial cells, 
$\alpha \approx 0.3$ \cite{hell}. The interbilayer coupling in the symmetric ternary system, makes the tagged SM movement slower, 
as seen by the lower value of $\alpha \approx 0.35$ (see Figure 5(c)). The tagged particle dynamics of the other lipids in the 
asymmetric bilayer, viz., POPC and cholesterol, show interesting differences between the two leaflets -- typically, the 
upper leaflet lipids show a smaller $\alpha$ ($\alpha = 0.45, 0.41$, respectively), than the lower leaflet lipids  
($\alpha = 0.85, 0.89$, respectively). 

\section*{Conclusion}

The cell membrane exhibits both lateral and transverse heterogeneity. In this paper, the equilibrium properties of a 
ternary component asymmetric bilayer membrane system at $l_o$-$l_d$ phase coexistence are studied using an atomistic MD simulation over 
a time scale of $220$\,ns. The asymmetric bilayer considered in this study is composed of  POPC, SM and cholesterol in the ratio of  
1:1:1 in the upper leaflet and POPC and cholesterol in the lower leaflet. The two significant results
from this study are: (i) cholesterol prefers to be associated with SM-rich domains ($l_o$-like) three times more that POPC-rich 
domains ($l_d$-like) and (ii) the saturated lipid SM in the $l_o$-like domain exhibits long-range tilt correlations purely as a result of the 
asymmetry in the bilayer composition (in contrast, the SM lipid molecules in the symmetric bilayer show no such tilt).

This bilayer-asymmetry induced lipid tilt in the $l_o$-like domain has important implications to local membrane deformation and 
hence membrane budding and endocytosis. 
%As suggested in many studies\cite{sarasij}, the existence of a lipid tilt expressed 
%over large scales provides a natural coupling to the local curvature tensor, and results in anisotropic bending stresses at the 
%membrane.
The existence of a lipid tilt expressed over large scales provides a natural coupling to the local curvature tensor, and results in anisotropic bending stresses at the membrane \cite{sarasij}.
 Moreover, if the constituent lipids are chiral (as they are in `raft'-lipids), then there would be additional bending 
stresses serving to deform the membrane locally. When the strength of these couplings are large enough, they can induce the local 
formation of  spherical buds or cylindrical  tubules. Recent FRET-based studies of the organization of lipid tethered proteins on 
the outer surface of living cells, such as GPI-anchored proteins, have shown that they form cholesterol sensitive
nanoclusters mediated by the activity of cortical actin \cite{sharma,debanjan}. These studies imply that there must exist a 
molecular linkage between the outer leaflet GPI-anchored proteins and cortical actin. The current study will form the basis for 
further investigations on possible transbilayer interactions between GPI-anchored proteins, SM and cholesterol, with specific 
saturated, long chain lipids at the inner leaflet that have potential interactions with actin or actin remodeling proteins.

%%%%%%%%%%%%%%%%%%%%%%%%%%%%%%%%%%%%%%%%%%%%%%%%%%%%%%%%%%%%%%%%%%%%%
%% The "Acknowledgement" section can be given in all manuscript
%% classes.  Rather than use \section, an appropriate macro is
%% provided that will always work.
%%%%%%%%%%%%%%%%%%%%%%%%%%%%%%%%%%%%%%%%%%%%%%%%%%%%%%%%%%%%%%%%%%%%%
\acknowledgement

We are very grateful to  M. Karttunen for generous advice and help on the use of the simulation packages without which this work would not have been possible. We thank  R. Sowdhamini for the use of computer facilities, and S. Mayor and V. A. Raghunathan for a critical reading of the manuscript. MR acknowledges  a grant from HFSP and CEFIPRA 3504-2.

%%%%%%%%%%%%%%%%%%%%%%%%%%%%%%%%%%%%%%%%%%%%%%%%%%%%%%%%%%%%%%%%%%%%%
%% The same is true for Supporting Information, which should use the
%% \suppinfo macro.
%%%%%%%%%%%%%%%%%%%%%%%%%%%%%%%%%%%%%%%%%%%%%%%%%%%%%%%%%%%%%%%%%%%%%
%\suppinfo
%
%The entire \textsf{achemso} bundle is generated by running
%\texttt{achemso.dtx} through \TeX. Running \LaTeX\ on the same file
%will generate the general documentation for both the class and
%package files.

%%%%%%%%%%%%%%%%%%%%%%%%%%%%%%%%%%%%%%%%%%%%%%%%%%%%%%%%%%%%%%%%%%%%%
%% The appropriate \bibliography command should be placed here.
%% Notice that the class file automatically sets \bibliographystyle
%% and also names the section correctly.
%%%%%%%%%%%%%%%%%%%%%%%%%%%%%%%%%%%%%%%%%%%%%%%%%%%%%%%%%%%%%%%%%%%%%
\bibliography{achemso}

\begin{figure*}[h!t]
\begin{center}
\includegraphics[angle=0,scale=0.5]{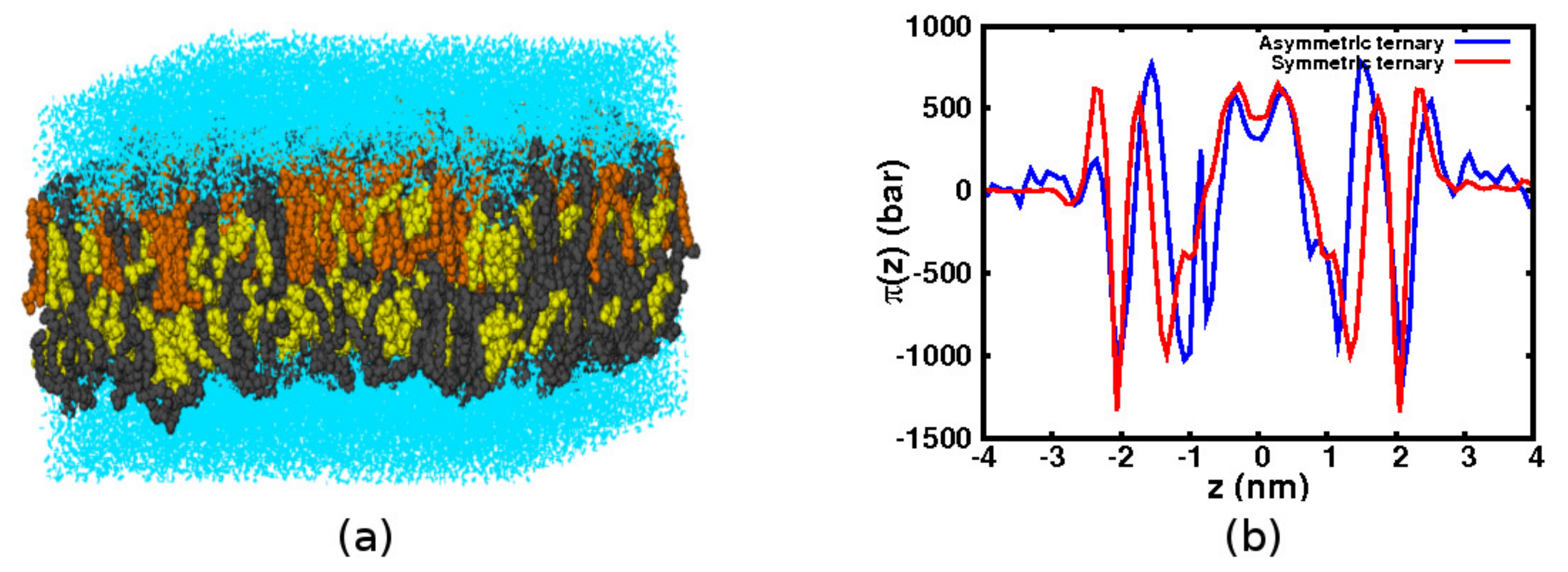}
\caption{(a) Snapshot (side view) at the end of the simulation of the ternary system POPC  (gray), SM (orange), Chol (yellow) forming the stable asymmetric bilayer
in water (cyan). (b) Lateral pressure profiles $\pi(z)$ as a function of position across the bilayer, for the ternary asymmetric bilayer (blue), and the ternary symmetric 
bilayer (red).
%due to the high elastic modulus of the lower leaflet as a consequence of high rigidity in the bilayer due to presence of high 
%concentration of cholesterol as seen in earlier simulations as well \cite{patra}.
}
\label{MD_snap}
\end{center}
\end{figure*}

\newpage

\begin{figure*}[h!t]
\begin{center}
\includegraphics[angle=0,scale=0.5]{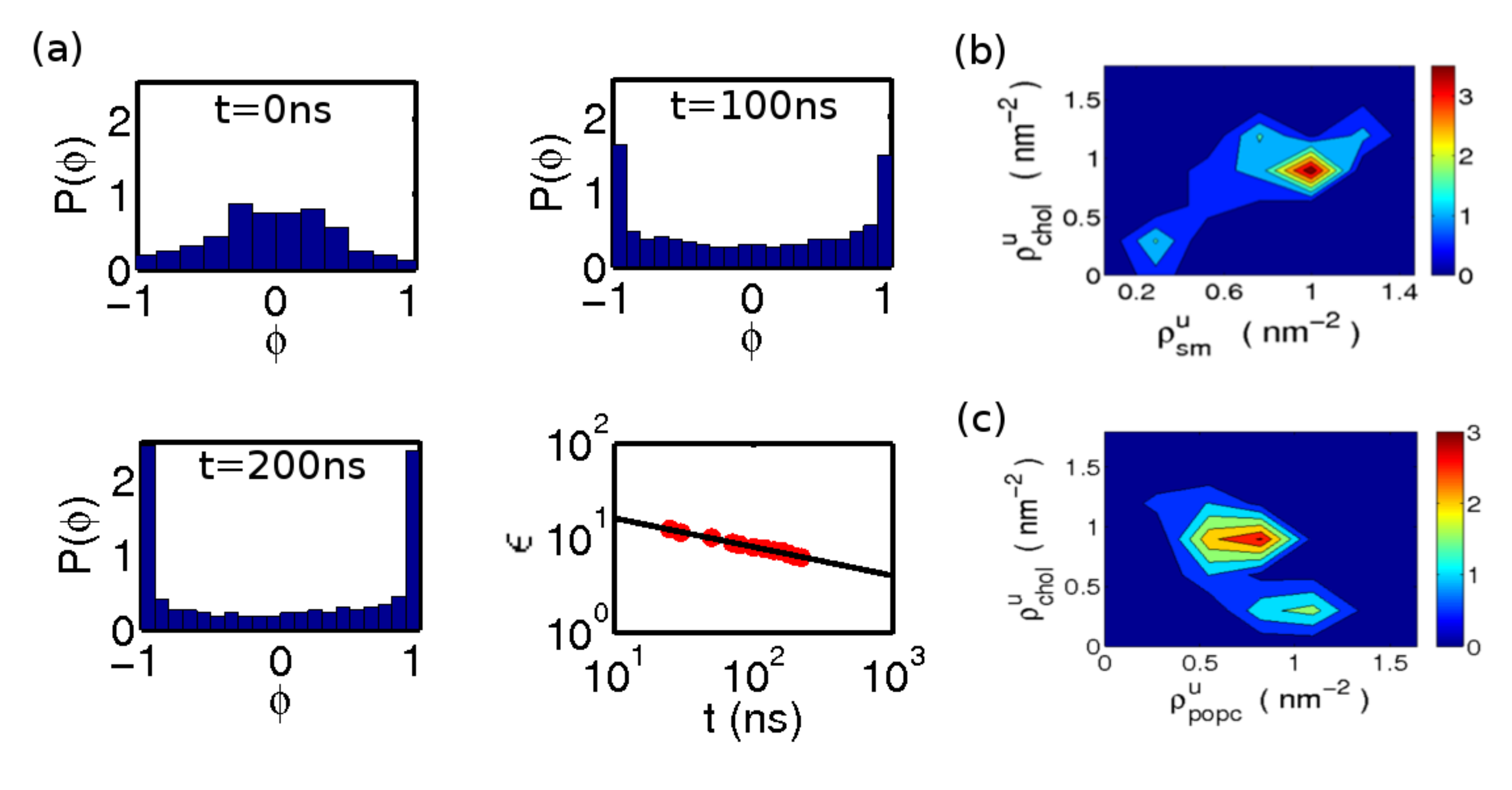}
\caption{(a) Probability distribution $P(\phi)$ (normalized) of the order parameter $\phi=\frac{\rho_{\tiny{\mbox{SM}}}-\rho_{\tiny{\mbox{POPC}}}}{\rho_{\tiny{\mbox{SM}}}+\rho_{\tiny{\mbox{POPC}}}}$, where $\rho_{\tiny{\mbox{SM}}}$ and $\rho_{\tiny{\mbox{POPC}}}$ are the concentrations of SM and POPC, respectively. Its time dependence from $t=0 - 200$\,ns, shows that the system, initially prepared in the mixed state ($\phi=0$), has coarsened 
into SM rich ($\phi=1$) and POPC rich ($\phi=-1$) domains separated by sharp interfaces. The last panel shows the dependence of the energy 
density $\epsilon$ with time. The data (red filled circles) suggests that $\epsilon$ goes as a power-law, $\epsilon \sim t^{-.3\pm0.15}$ (fitted line), with an exponent consistent with dynamical scaling, $z=3$ (see text). (b)
Joint probability distribution (color bar) of the concentration (in units of  $\rm{number}/nm^2$) of (b) SM and Chol and (c) POPC and Chol, coarse-grained over $[1.73$\,nm$]^2$
in the upper leaflet and averaged over $20$\,ns. Red shows the highest joint probability and blue the lowest. Note the 
 strong correlation (anticorrelation) between SM-Chol (POPC-Chol), respectively. The figures clearly show an enrichment of cholesterol in SM-enriched domains by 
 a ratio $3:1$}
\label{N_p-s-c}
\end{center}
\end{figure*}

\newpage

\begin{figure*}[h!t]
\begin{center}
\includegraphics[angle=0,scale=0.5]{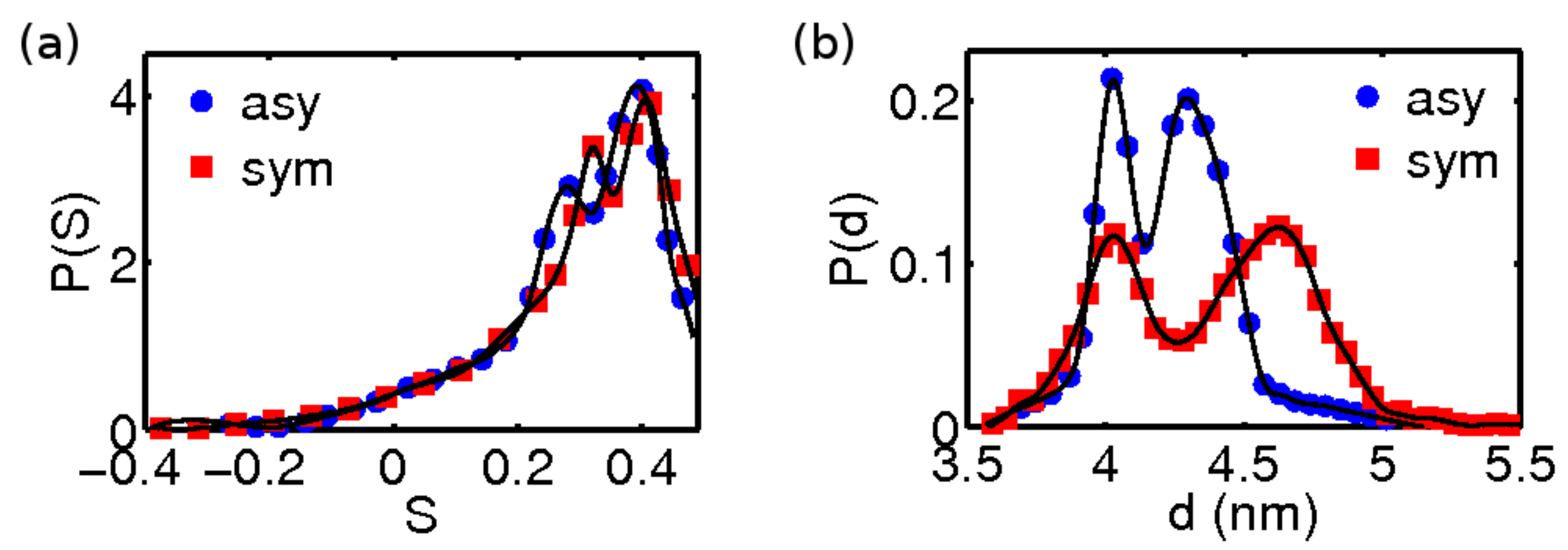}
\caption{(a) Probability distribution of the deuterium order parameter $S$  in the asymmetric (blue) and symmetric (red) bilayers. 
This is then coarse-grained over a spatial scale of $1.56$\,nm and time scale of $20$\,ns. Here $S$ of the selected carbons (C5-C7) of 
POPC and SM is displayed, after suitable binning in the xy-plane. with the lower branch $0.1-0.26$ corresponding to the low 
partitioning of SM in the $l_d$ phase and the higher branch $0.35-0.45$ corresponding to its enrichment in the $l_o$ phase.
(b) Probability distribution of the bilayer thickness $d$  in the asymmetric (blue) and symmetric (red) bilayers. The two distinct 
peaks in the asymmetric bilayer at $4$\,nm and $4.3$\,nm indicates the coexistence of the
$l_d$ and $l_o$ phases.  The local bilayer thickness is constructed by via x-y grids  for the upper and lower
 leaflet of the bilayer, with grid size $=1.95$\,nm. An average over the z-coordinates of the phosphorus atom in the head groups 
of POPC and SM within each x-y grid is then performed. The difference between the z-coordinates corresponding to the same 
x-y grid of the two leaflets gives the local bilayer thickness.
}
\label{th_PSC}
\end{center}
\end{figure*}

\newpage

\begin{figure*}[h!t]
\begin{center}
\includegraphics[angle=0,scale=0.4]{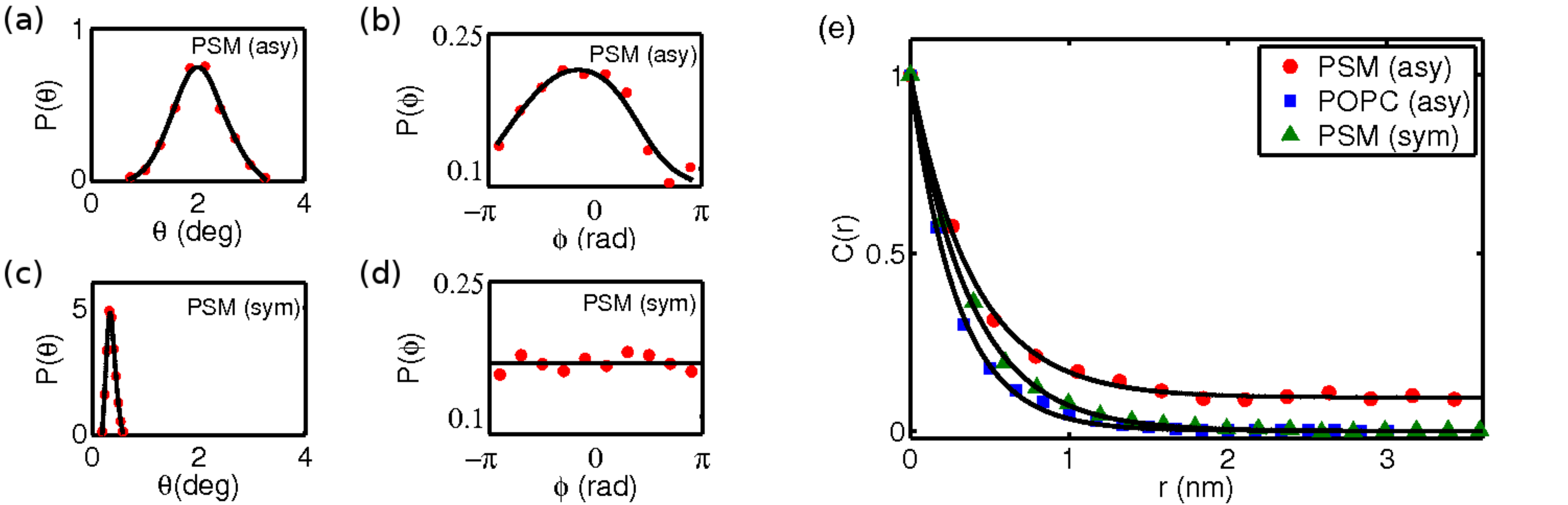}
\caption{Probability distribution of the coarse-grained tilt angles $\theta$ and $\phi$ of SM  (see text) in (a, b) the asymmetric bilayer, and (c, d) symmetric bilayer, respectively. 
These data are collected from equilibrium configurations that exhibit complete phase segregation, where the domain size is about half the system size.
In the asymmetric ternary bilayer, SM exhibits a measurable tilt, as evidenced from the distributions of $\theta$ and $\phi$. In comparison, SM in the symmetric bilayer shows an absence of tilt -- the distribution of $\theta$ is peaked at a value close to $0$, while the distribution of $\phi$ is uniform. (e) The tilt correlation function $C(r) \equiv \langle \phi(r) \phi(0)\rangle$, normalized to its value at $r=0$, shows
an exponential decay to zero for POPC (asymmetric bilayer) and SM (symmetric bilayer). However SM in the asymmetric bilayer shows an exponential
decay to a {\it nonzero value}, demonstrating long range tilt correlations. The correlation length, defined by the scale at which $C(r)$ decreases to $1/e$ of its value at $r=0$, can be easily read out,  $\xi_{\tiny{\mbox{SM}}}(sym)=0.35\,nm$, $\xi_{\tiny{\mbox{SM}}}(asy)=0.5\,nm$ and $\xi_{\tiny{\mbox{POPC}}}(asy)= 0.3\,nm$.
}
\label{tilt_P_PC}
\end{center}
\end{figure*}

\newpage

\begin{figure*}[h!t]
\begin{center}
\includegraphics[angle=0,scale=0.4]{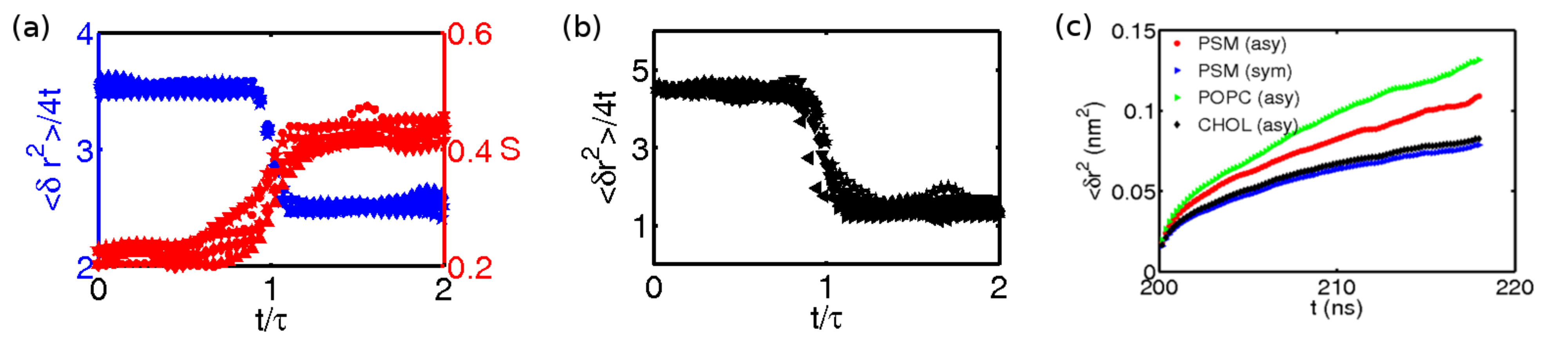}
\caption{Mean square displacement of tagged particles as a function of time collected at high resolution over short time scales. Tagged particle diffusion of (a) SM and (b) Chol in the asymmetric bilayer-- statistics collected over $6$ tagged particles, each starting from the $l_d$ domain, shows a data collapse onto a nonlinear crossover scaling curve. The crossover from high to low diffusion
occurs at a time $\tau$, computed as the first-passage time of the tagged particle.  For SM, we find that  $D_0 = 3.49 \mu$m$^2$s$^{-1}$ and 
$D_{\infty}=2.69 \mu$m$^2$s$^{-1}$, while for Chol, $D_0 = 4.23 \mu$m$^2$s$^{-1}$ and 
$D_{\infty}=1.36 \mu$m$^2$s$^{-1}$.
Note that in (a) as SM moves across domains, the deuterium order parameter $S$ of tagged SM shows a similar crossover, going from $\sim 0.22$ ($l_d$) to $\sim 0.42$ ($l_o$). (c) MSD versus time with fits to $\langle \delta r_i(t)^2 \rangle \propto t^{\alpha}$, for SM in the asymmetric ($\alpha = 0.39$) and symmetric bilayers ($\alpha = 0.35$), and POPC and Chol in the upper leaflet of the asymmetric bilayer ($\alpha = 0.45$ and $0.41$, respectively). Note that the $\alpha$ exponent for SM in the symmetric bilayer is smaller than the asymmetric bilayer, due to the transbilayer coupling.  Likewise, the MSD for POPC and Chol in the symmetric ternary bilayer are shown in Supplementary Figure S10, for comparison.}
\label{crossover}
\end{center}
\end{figure*}

%%%%%%%%%%%%%%%%%%%%%%%%%%%%%%%%%%%%%%%% Supplementary Info %%%%%%%%%%%%%
\newpage

\vskip 5cm

\begin{center}
{\large{\bf Supplementary Information\\
A. Polley et al.}}
\end{center}

%\title{Supplementary Information\\
%A. Polley et al.}

%\maketitle

\begin{center}
{\large{\bf Tabulation of force, torque and tension in model bilayers at equilibrium}}
\end{center}

\begin{enumerate}

\item {\bf Components of Force}

{\underline {POPC (symmetric)}}\\
 $F_{1}=0.83 \pm 1.66$\,nN, $F_{2}=0.27\pm1.14$\,nN, $F_{3}=-0.16\pm1.25$\,nN

{\underline{POPC-CHOL (symmetric)}}\\
$F_{1}=-1.01\pm1.36$\,nN, $F_{2}=-0.32\pm2.72$\,nN, $F_{3}=0.18\pm0.62$\,nN

\item {\bf Components of the moment of force}

{\underline {POPC (symmetric)}}\\
 $M_{12}=8.26\pm0.102$\,nN$\cdot$nm, $M_{13}=10.74\pm0.066$\,nN$\cdot$nm, $M_{23}=9.33\pm0.094$\,nN$\cdot$nm

{\underline{POPC-CHOL (symmetric)}}\\
$M_{12}=5.94\pm0.082$\,nN$\cdot$nm, $M_{13}=8.882\pm0.23$\,nN$\cdot$nm, $M_{23}=10.757\pm0.42$\,nN$\cdot$nm

\item {\bf Surface Tension}

{\underline {POPC (symmetric)}}\\
$\gamma=0.0204\pm0.0357$\,bar-nm

{\underline{POPC-CHOL (symmetric)}}\\
$\gamma=0.0054\pm0.0099$\,bar-nm

The values of the components of force, moment of force and surface tension in these symmetric 1 and 2 component systems are comparable to those
of the asymmetric ternary system given in the main text. This implies that the asymmetric ternary bilayer is as mechanically stable as the symmetric systems.

\end{enumerate}

%%%%%%%%%%%%%%%%%%%%%%%%%%%%%%%%%%%%%%%% Supplementary Figure %%%%%%%%%%%%%
\newpage

\vskip 5cm

\begin{center}
{\Huge{\bf Supplementary Figures\\
A. Polley et al.}}
\end{center}

\vskip 5cm

\renewcommand{\thefigure}{\arabic{figure}}
\addtocounter{figure}{-5}

\begin{figure*}[h!t]
\begin{center}
\includegraphics[angle=0,scale=0.6]{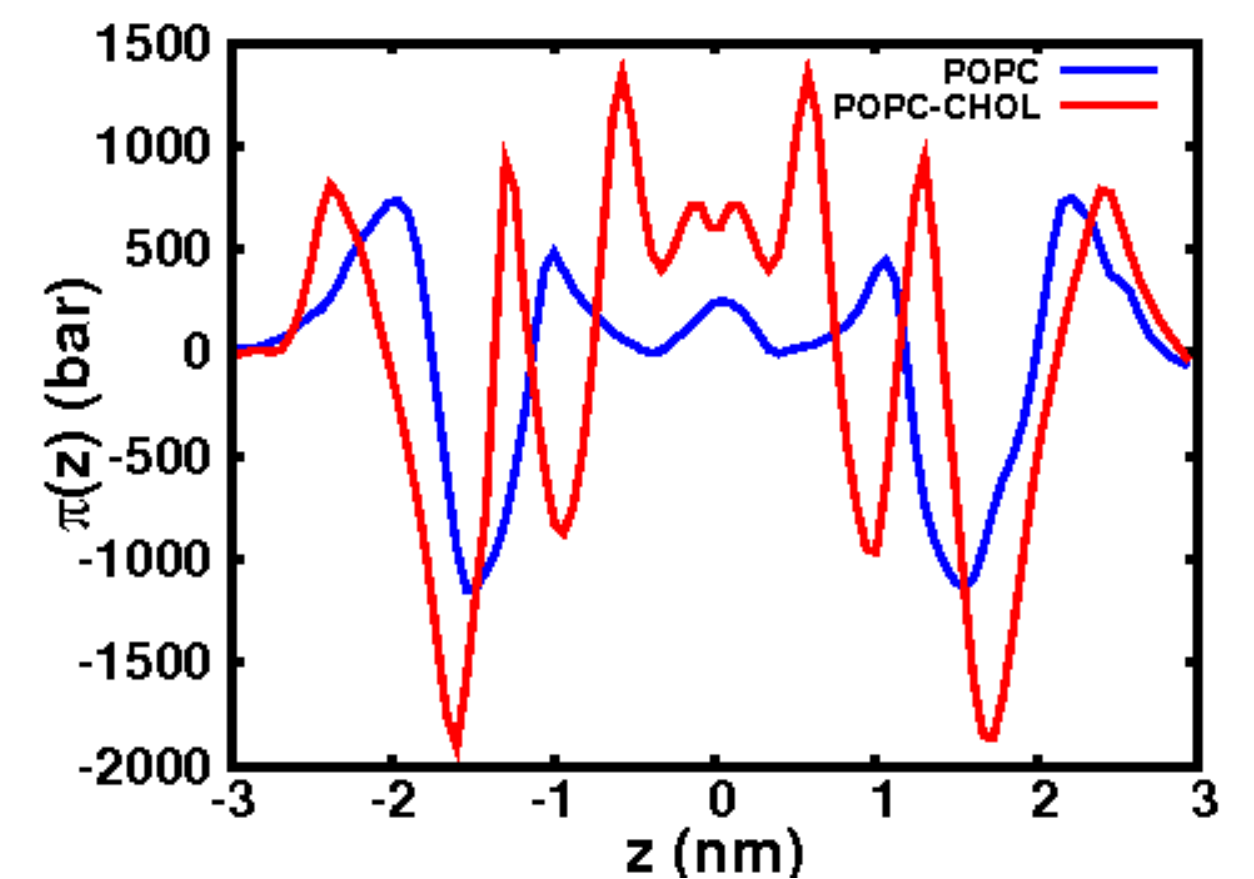}
\caption{Lateral pressure profile of the symmetric bilayer made of (a) pure POPC (blue) and (b) POPC + Chol, with ratio $1:1$ (red). }
\label{S1}
\end{center}
\end{figure*}

\newpage

\vskip 5cm

\begin{figure*}[h!t]
\begin{center}
\includegraphics[angle=0,scale=0.275]{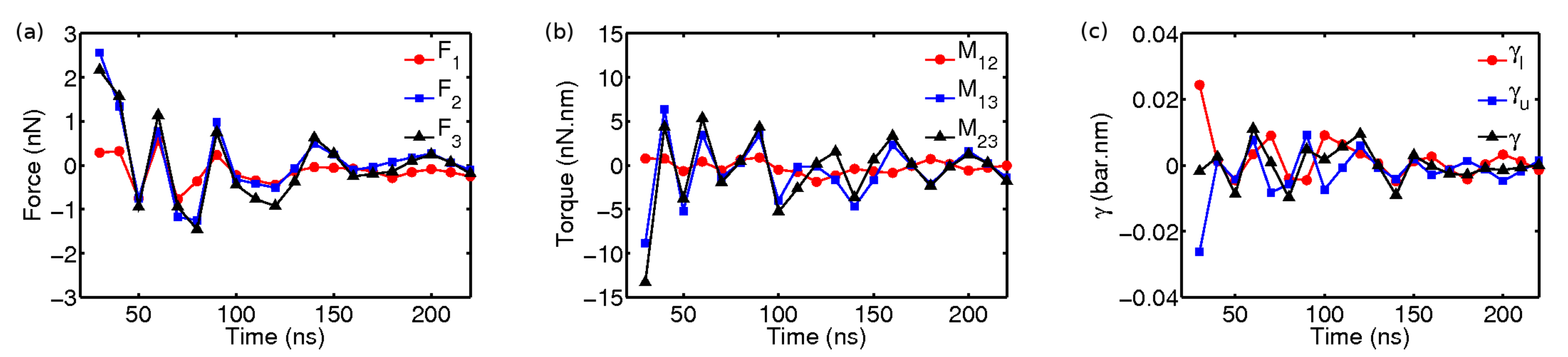}
\caption{(a) Time dependence of the components of the force (defined in main text), where each data point  is an average over $10$\,ns .  (b) Time dependence of the components of the torque (defined in main text) where each data point is averaged over $10$\,ns. The observation that at late times these components fluctuate about zero,
and that these fluctuations are not large, implies that in the mean there is force and torque balance. (c) The time dependence of the surface tension (evaluated from the components of the stress tensor) in the upper ($\gamma_{u}$) and lower ($\gamma_{l}$) leaflets of the membrane and the total surface tension ($\gamma$). The observation that both $\gamma_u$ and $\gamma_l$  are small and fluctuate about zero, implies that in the mean there is no imbalance of surface tension in the two
leaflets. The observation that the total $\gamma$ is small and fluctuates about zero implies that in the mean the bilayer membrane is tensionless.
}
\label{S2}
\end{center}
\end{figure*}

\newpage

\vskip 5cm

\begin{figure*}[h!t]
\begin{center}
\includegraphics[angle=0,scale=0.55]{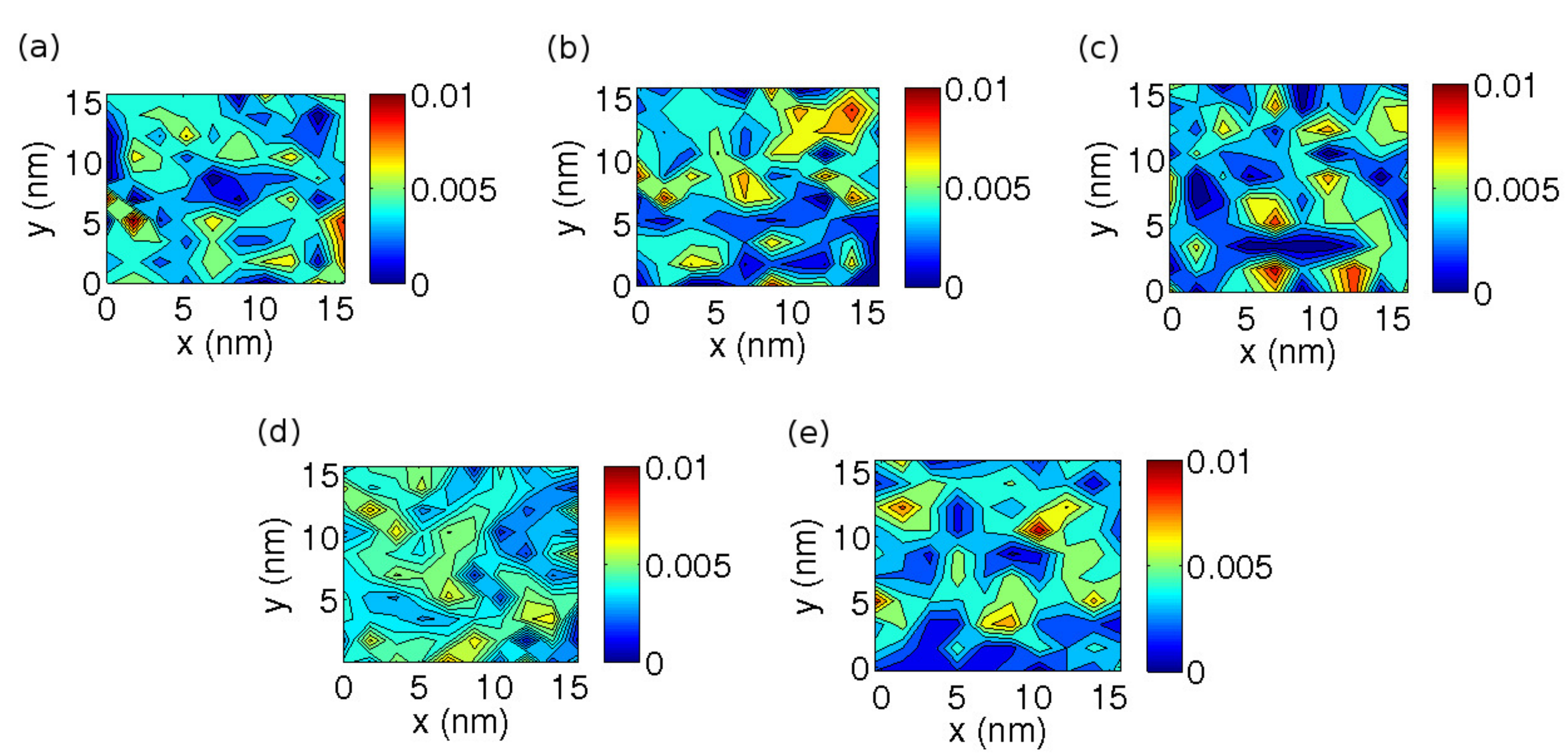}
\caption{Spatial variation of lipid concentration in the asymmetric ternary bilayer system : (a)
POPC, (b) PSM, (c) Chol in the upper leaflet, and (d) POPC and (e)
Chol in the lower leaflet.  The LUT bars denote the fraction of the lipid species within an area $[1.56\,nm]^2$. }
\label{S3}
\end{center}
\end{figure*}

\newpage

\vskip 5cm

\begin{figure*}[h!t]
\begin{center}
\includegraphics[angle=0,scale=0.5]{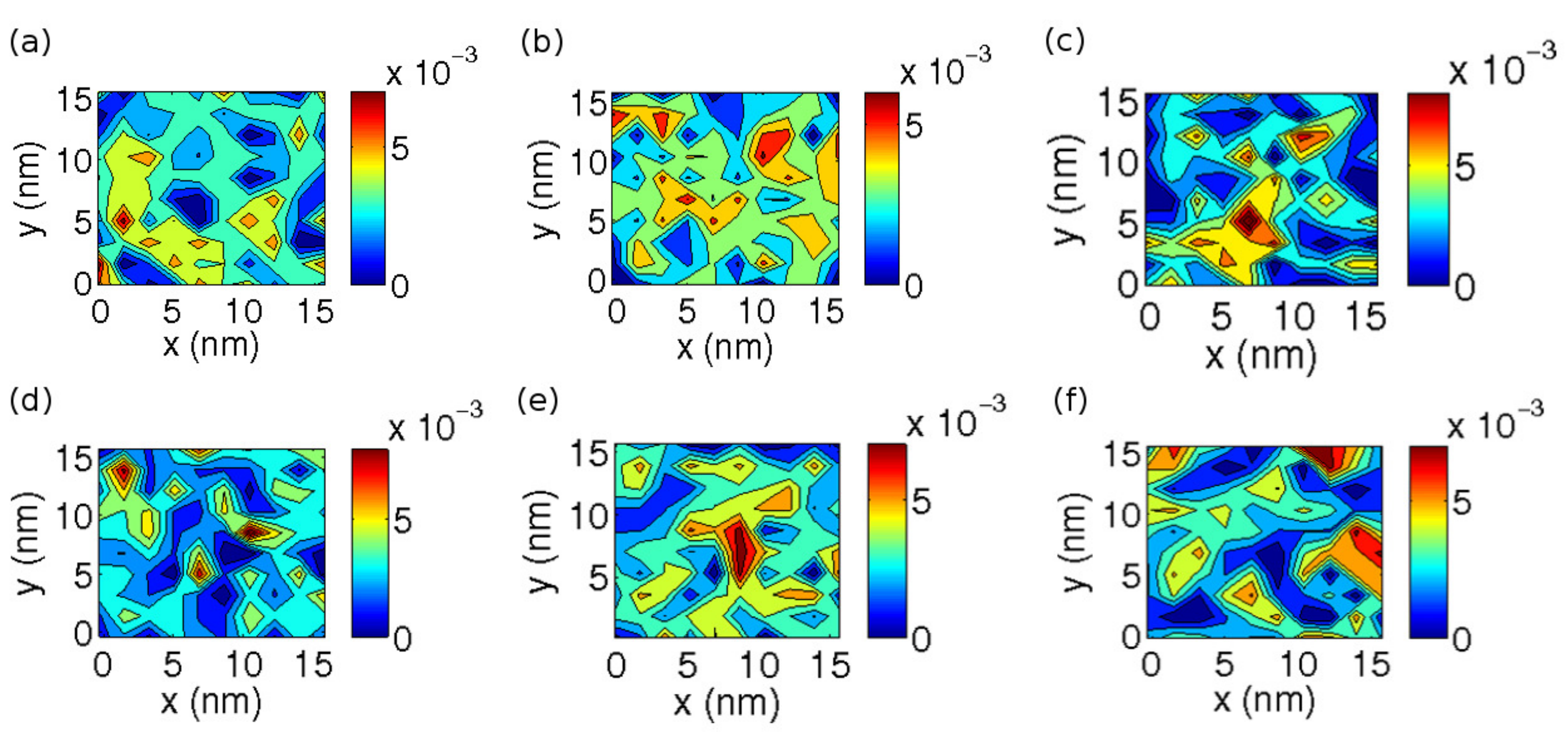}
\caption{Spatial variation of lipid concentration in the symmetric ternary bilayer system : (a)
POPC, (b) PSM, (c) Chol in the upper leaflet, and (d) POPC, (e) PSM
and (f) Chol in the lower leaflet. The LUT bars denote the fraction of the lipid species within an area $[1.56\,nm]^2$. Note from (b) that the domain sizes are larger than in the
asymmetric bilayer. Further, from (b) and (e), we note that there is significant registry of the domains across the bilayer (see Fig. 5).}
\label{S4}
\end{center}
\end{figure*}

\newpage

\vskip 5cm

\begin{figure*}[h!t]
\begin{center}
\includegraphics[angle=0,scale=0.45]{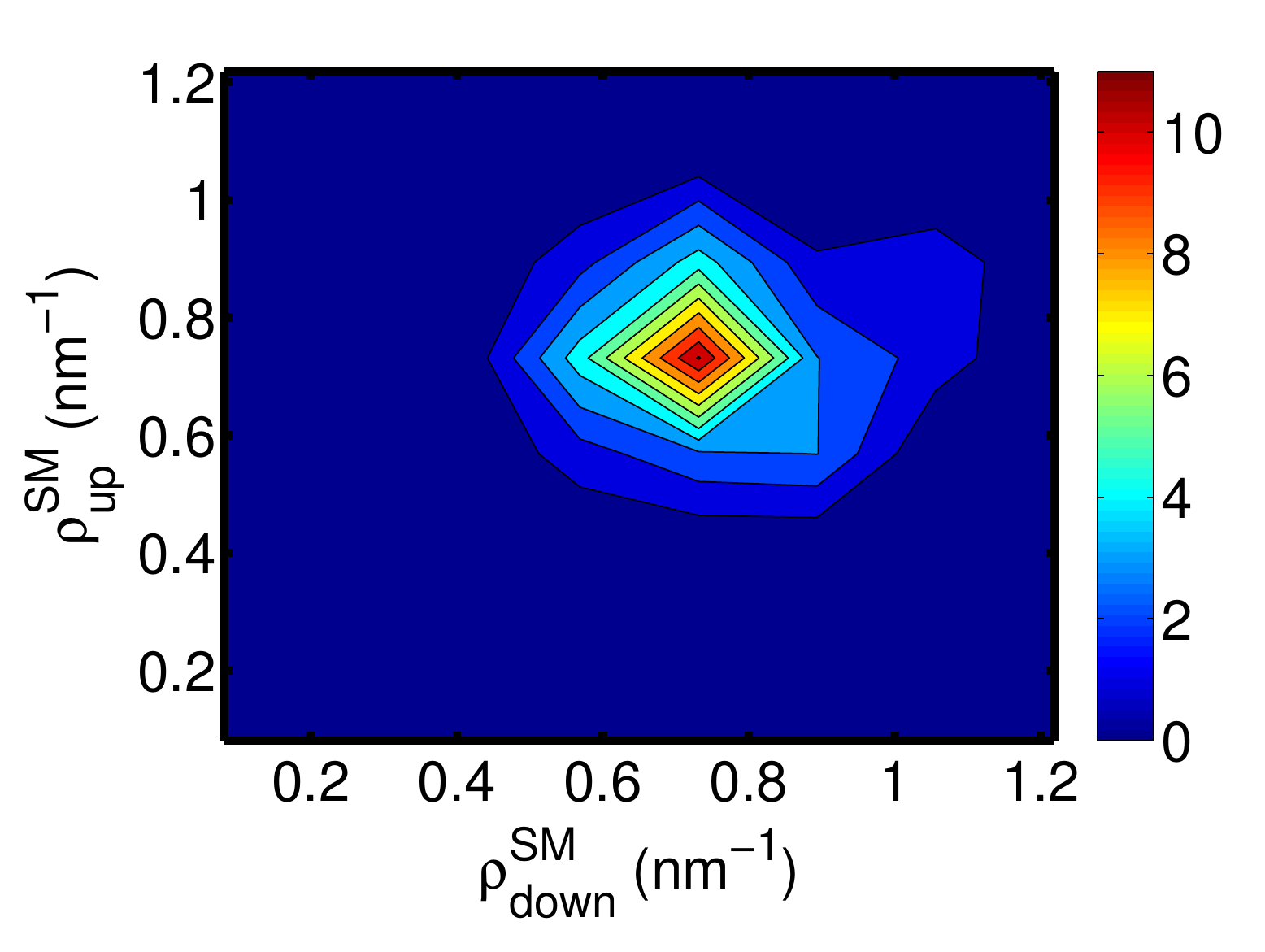}
\caption{Joint probability distribution (color bar) of the concentration (in units
of $number/nm^2$) of SM of upper leaflet and SM at lower leaflet of symmetric ternary bilayer coarse-grained over $[1.73 \,nm]^2$ and 
averaged over 20 ns. Red shows the highest joint probability and blue the
lowest. Note the strong correlation between SM-SM in the upper and lower leaflets implies significant bilayer registry in the SM-enriched domain across the two leaflets. }
\label{S5}
\end{center}
\end{figure*}

\newpage

\vskip 5cm

\begin{figure*}[h!t]
\begin{center}
\includegraphics[angle=0,scale=0.275]{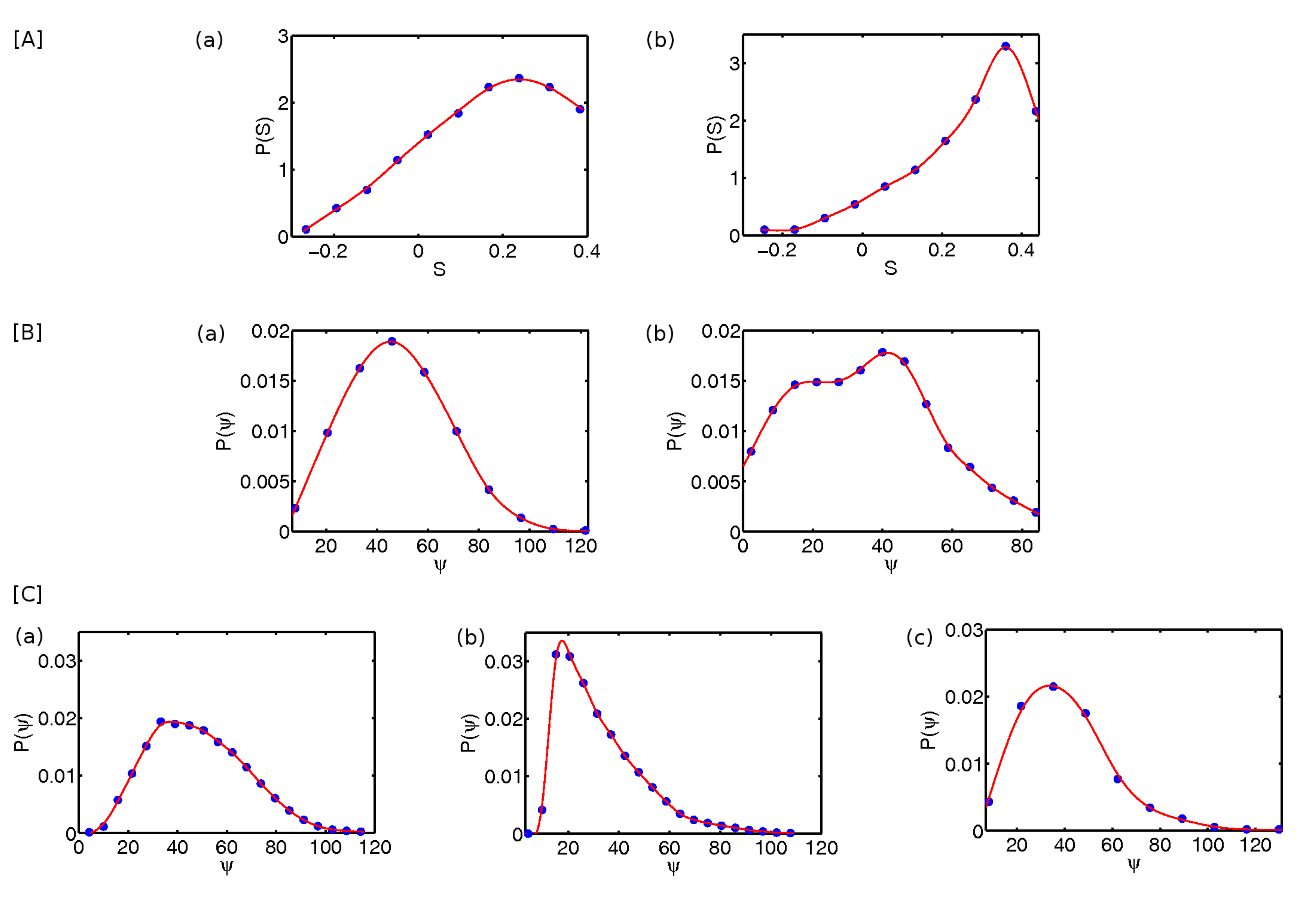}
\caption{[A] Probability distribution of the deuterium order parameter $S$ of POPC in the symmetric bilayer  of  (a) POPC and (b) POPC + CHOL (with
ratio $1:1$). [B]
Probability distribution of splay of POPC in the symmetric bilayer of (a) POPC and (b) POPC-CHOL (with ratio $1:1$).
[C] Probability distribution of splay of (a) POPC and (b) PSM  in the upper leaflet and (c)  POPC in the  lower  leaflet of the asymmetric ternary bilayer. }
\label{S6}
\end{center}
\end{figure*}

\newpage

\vskip 5cm

\begin{figure*}[h!t]
\begin{center}
\includegraphics[angle=0,scale=0.3]{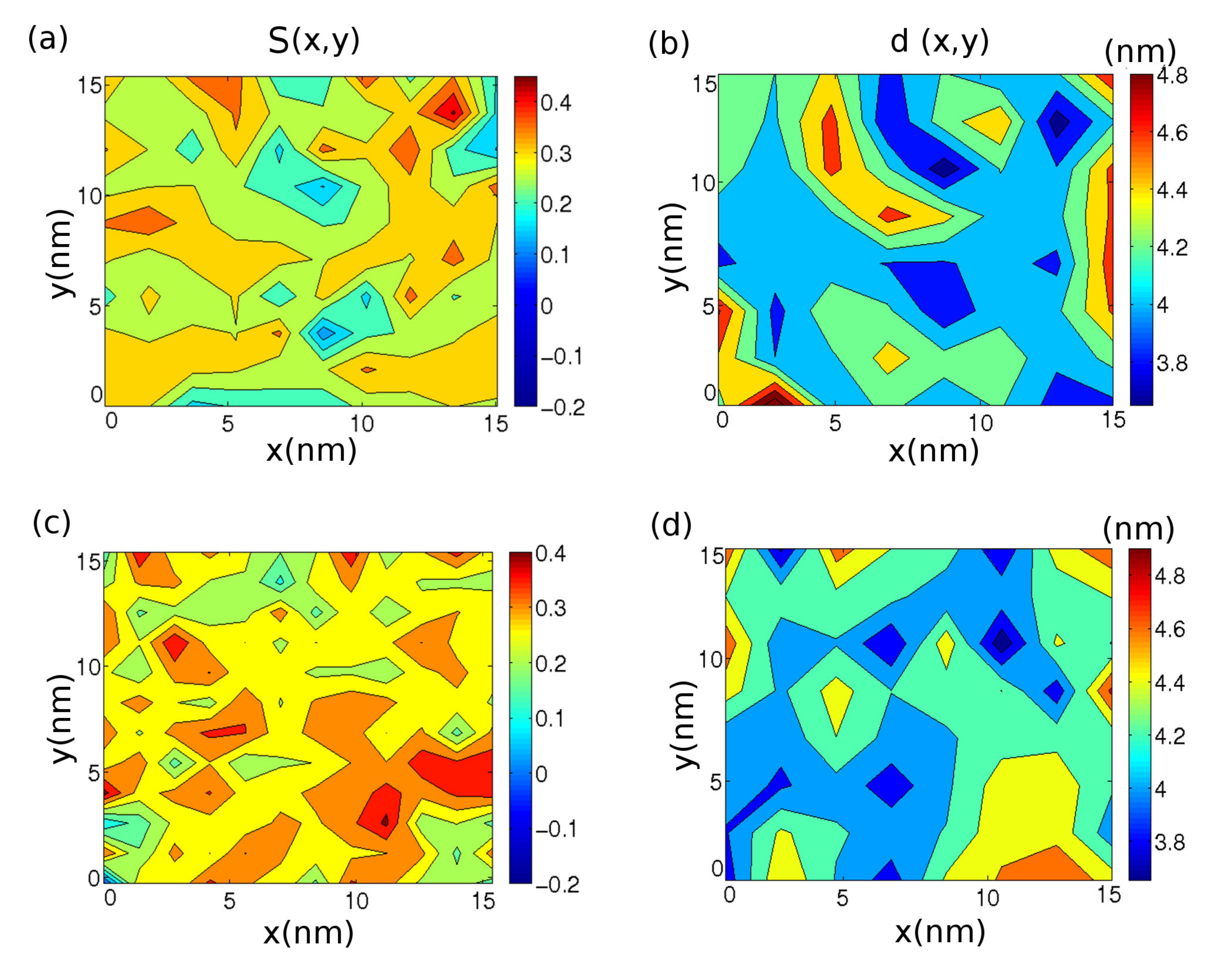}
\caption{Spatial variation of deuterium order parameter $S(x,y)$ and bilayer thickness $d(x,y)$ of (a-b) asymmetric bilayer and (c-d) symmetric bilayer of the ternary system. 
In both cases, the domains coarsen over time; note however that the domain are larger in the symmetric bilayer compared to the asymmetric bilayer, evaluated over the same time. In addition, note that in the case of the symmetric bilayer, the
domain registry across the bilayer is significant, consistent with Fig. 5.}
\label{S7}
\end{center}
\end{figure*}

\newpage

\vskip 5cm

\begin{figure*}[h!t]
\begin{center}
\includegraphics[angle=0,scale=0.4]{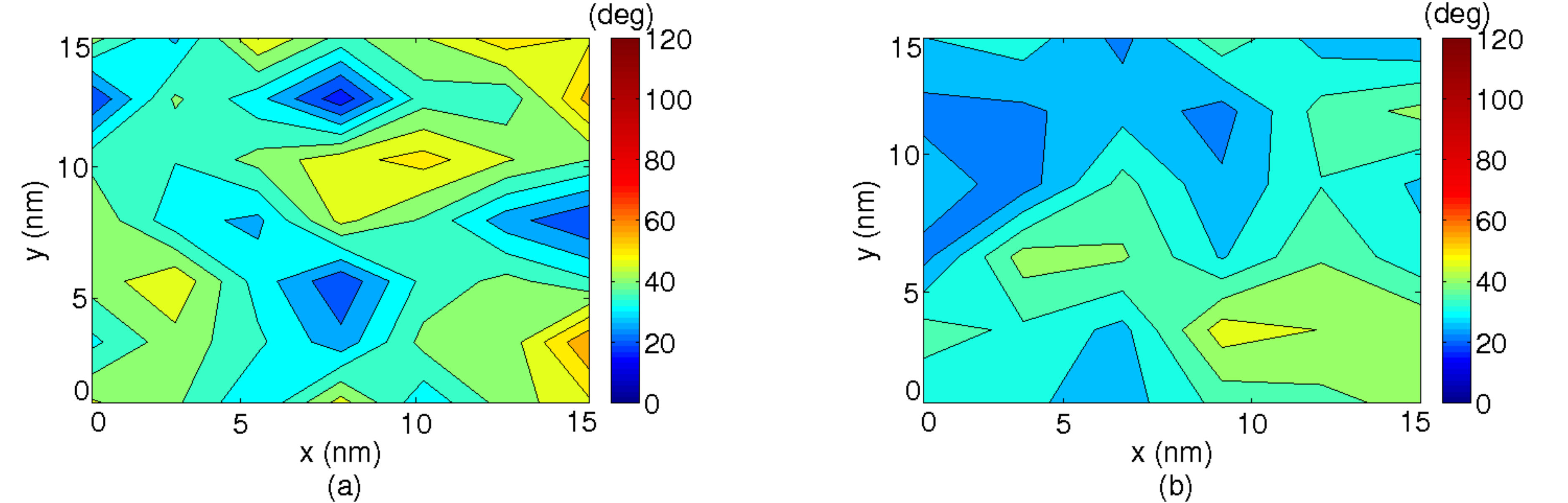}
\caption{Spatial heterogeneity of splay of (a) POPC and (b) PSM in the upper leaflet of the asymmetric ternary bilayer.}
\label{S8}
\end{center}
\end{figure*}

\newpage

\vskip 5cm

\begin{figure*}[h!t]
\begin{center}
\includegraphics[angle=0,scale=0.8]{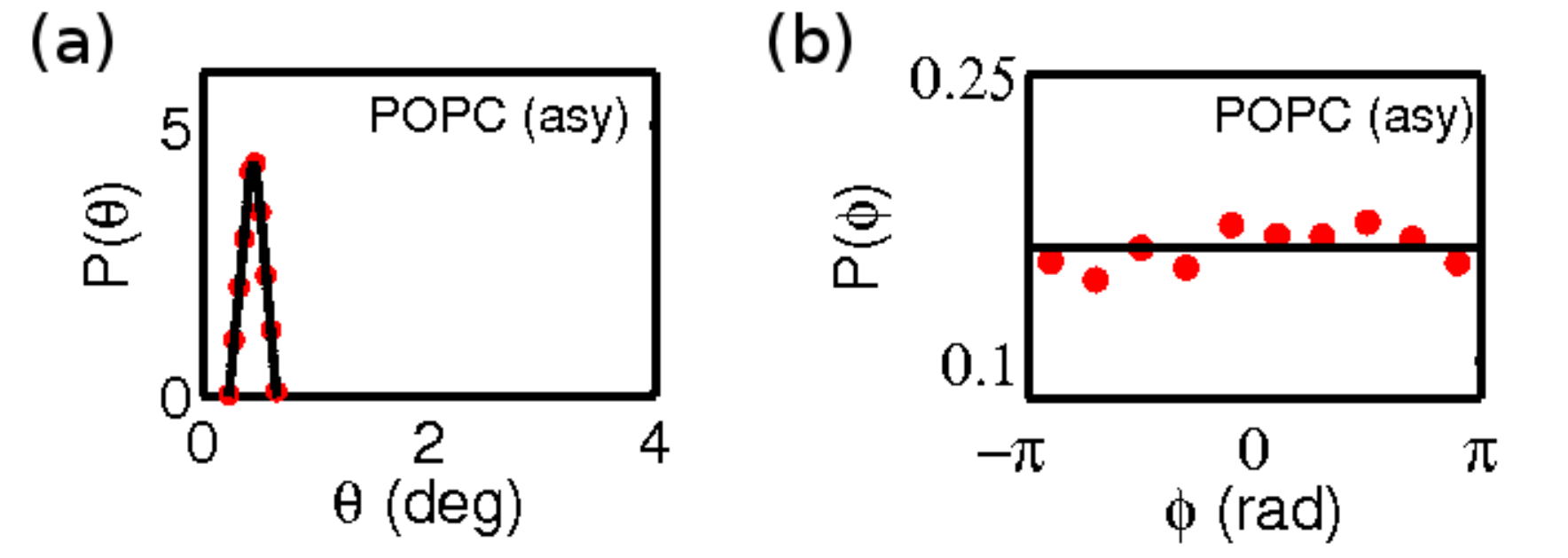}
\caption{Probability distribution of the tilt angles (a) $\theta$ and (b) $\phi$ for POPC in the asymmetric ternary bilayer.
Note that the distribution of $\theta$ is peaked about  $\approx 0$, while the distribution of $\phi$ is uniform, consistent with
the known fact that POPC does not exhibit a tilt at these temperatures.}
\label{S9}
\end{center}
\end{figure*}

\newpage

\vskip 5cm

\begin{figure*}[h!t]
\begin{center}
\includegraphics[angle=0,scale=0.45]{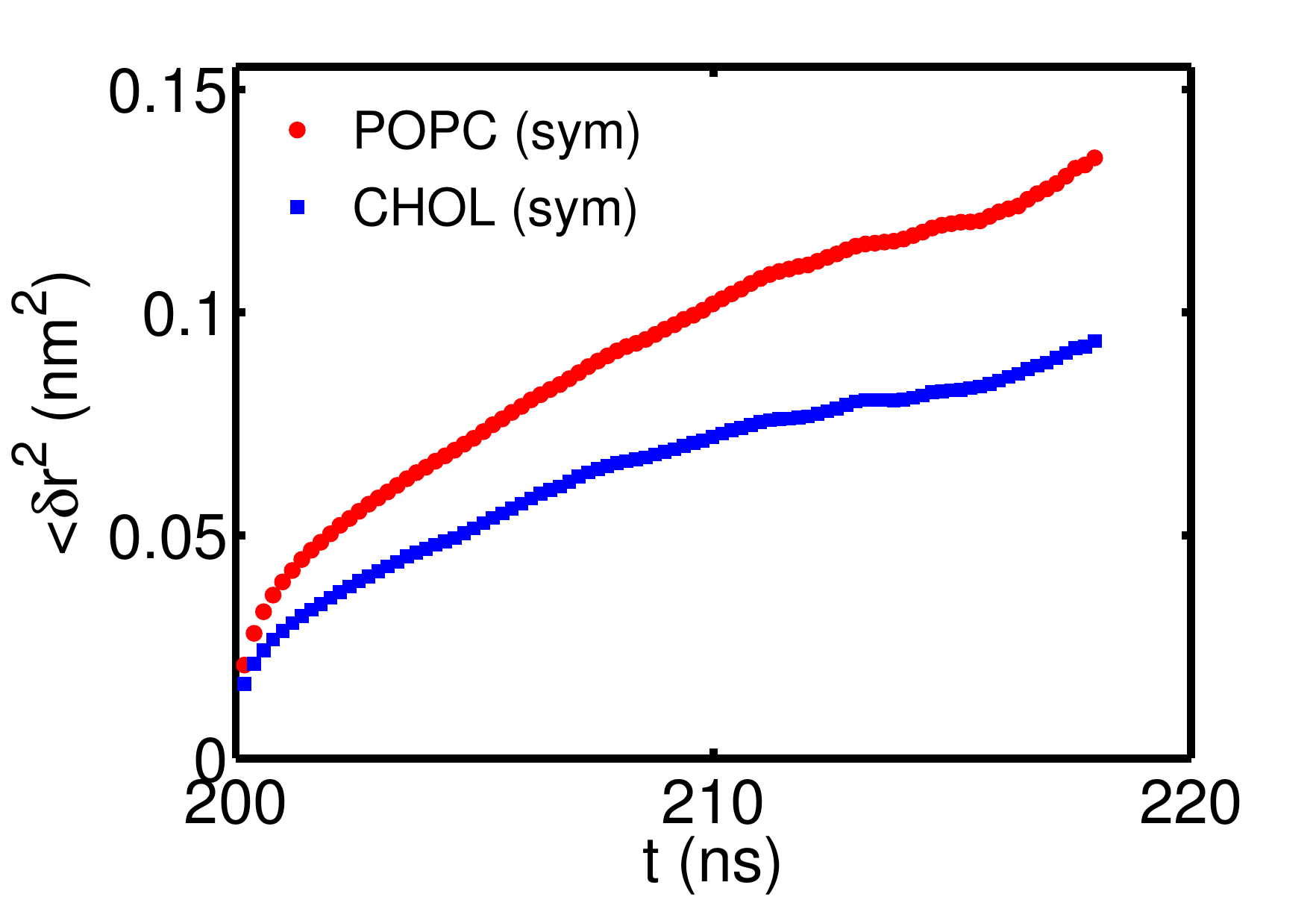}
\caption{MSD versus time for tagged POPC and Chol in the symmetric ternary bilayer. Fits to $\langle \delta r_i(t)^2 \rangle \propto t^{\alpha}$ gives $\alpha = 0.35$ (POPC) and 
$\alpha =0.32$ (Chol). Note that the values of $\alpha$ are consistently lower than those in the asymmetric ternary bilayer.}
\label{S10}
\end{center}
\end{figure*}

\end{document}